\documentclass[12pt,preprint]{aastex}

\shorttitle{Millimeter and Submillimeter Survey of the R~CrA Region}
\shortauthors{Groppi et al.}

\begin{document}

\title{Millimeter and Submillimeter Survey of the R Corona Australis Region}
\author{Christopher E. Groppi}
\affil{National Radio Astronomy Observatory, Tucson, AZ 85721}
\email{cgroppi@nrao.edu}
\author{Craig Kulesa and Christopher Walker}
\affil{Steward Observatory, University of Arizona, Tucson, AZ 85721}
\author{Christopher L. Martin}
\affil{Smithsonian Astrophysical Observatory, Cambridge, MA 02138}

\begin{abstract}
Using a combination of data from the Antarctic Submillimeter Telescope and Remote Observatory (AST/RO), the Arizona Radio Observatory Kitt Peak 12m telescope and the Arizona Radio Observatory 10m Heinrich Hertz Telescope, we have studied the most active part of the R~CrA molecular cloud in multiple transitions of Carbon Monoxide, HCO$^+$ and 870\micron\ continuum emission.  Since R~CrA is nearby (130 pc), we are able to obtain physical spatial resolution as high as 0.01pc over an area of 0.16 pc$^2$, with velocity resolution finer than 1 km/s. Mass estimates of the protostar driving the mm-wave emission derived from HCO$^+$, dust continuum emission and kinematic techniques point to a young, deeply embedded protostar of $\sim$0.5-0.75 M$_\odot$, with a gaseous envelope of similar mass. A molecular outflow is driven by this source that also contains at least 0.8 M$_\odot$ of molecular gas with $\sim$0.5 L$_\odot$ of mechanical luminosity. HCO$^+$ lines show the kinematic signature of infall motions as well as bulk rotation. The source is most likely a Class 0 protostellar object not yet visible at near-IR wavelengths. With the combination of spatial and spectral resolution in our data set, we are able to disentangle the effects of infall, rotation and outflow towards this young object.
\end{abstract}


\section{Introduction}

R Corona Australis (R CrA) is one of the most nearby active star forming regions, at $~\sim$130pc \citep{mar81}. This close proximity allows for high spatial resolution observations, even with single dish millimeter wave and submillimeter wave telescopes. The cloud is home to many Herbig Ae/Be and T Tauri stars that have been studied extensively in the near infrared \citep{wil97}. Recent observations in millimeter wave continuum \citep{hen94} and molecular lines (\cite{har93}, \cite{and97-1}, \cite{and97-2}), have shown the existence of active, embedded star formation not directly associated with R~CrA, but most likely associated with the Class I source IRS7 \citep{tay84}. IRS7 is the most deeply embedded near infrared point source in the immediate area.

R~CrA was first studied in CO by Loren \citep{lor79}. His fairly low spatial resolution maps (2.4\arcmin) showed large amounts of CO peaking at the general location of R~CrA, with high velocity wing emission throughout the map. The wing emission was later interpreted as a molecular outflow by \cite{lev88}. Levreault's higher spectral and spatial resolution data revealed a large bipolar outflow with an extent of over 10\arcmin. He also suggested that two outflows might be responsible for the observed morphology, and that the embedded source IRS7 might be responsible for the outflow, not R~CrA itself. 

In 1985, Wilking et al. used the Kuiper Airborne Observatory to map R~CrA in 100~\micron~ and 50~\micron~ continuum emission \citep{wil85}. The spatial resolution was not adequate to determine if R~CrA, IRS7 or some other unidentified source was driving the continuum and molecular line emission. Since R~CrA and IRS7 are both thought to be high mass sources visible in the near-IR, there is a possibility the driving source is neither of these objects. At both 50 and 100~\micron~, the emission was found to be extended, with a flux density of 570 Jy/beam at 100~\micron~. These FIR continuum observations were supplemented by Henning \citep{hen94}. This 23\arcsec ~ resolution map spatially resolved the region, and showed that the peak of the FIR emission is not associated with R~CrA, but is much closer to the infrared source IRS7.

In 1993, Harju et al. performed a large area (40\arcmin$\times$10\arcmin) survey of the entire R~CrA molecular cloud in C$^{18}$O \citep{har93}. They estimated that the entire cloud complex contains more than 120 M$_\sun$ of molecular gas. The densest part of this core, centered around IRS7 and R~CrA contains $\sim$60 M$_\sun$ of molecular gas. Their kinematic analysis showed evidence for  rotation around IRS7, and also clearly identified outflow lobes in the area. Follow-up work by Anderson \citep{and97-1} identified a large molecular disk around IRS7, which shows the kinematic signature of rotation. They mapped the region around R~CrA in multiple transitions of HCO$^+$ to trace the dense gas. In HCO$^+$, the outflow lobes have an extent of 4\arcmin, and show no evidence for the multiple outflows suggested by Levreault. In order to probe the disk dynamics further, the central 2\arcmin$\times$2\arcmin\ region of R~CrA was mapped in HCO$^+$(3-2). The data show an elongated structure, with three peaks to the northwest, southeast and south of IRS7. A slice in centroid velocity through IRS7 at a position angle of 45 degrees shows signs of rotation. Anderson then followed up their previous work with HCO$^+$(4-3) mapping of the central 2\arcmin$\times$2\arcmin ~ around IRS7, providing improved spatial resolution and kinematic information \citep{and97-2}. The integrated intensity map shows a structure elongated in the N-S direction, unresolved in the E-W direction, peaking on IRS7. The multi-lobed structure visible in HCO$^+$(3-2) is not present, which they explain as the effect of self-absorption on the line profiles. Velocity centroid maps show a strong velocity gradient near IRS7, and a slice orthogonal to the gradient at a position angle of about 45 degrees revealed the characteristic signature of rotational motion. They fit this rotation curve with a disk+central point source model and derived a central mass of 0.8 M$_\sun$. This result points towards the conclusion that the mm-wave source is a deeply embedded protostar of fairly low mass. 

\section{Observations}

Observations for this study were performed with the 1.7m Antarctic Submillimeter Telescope and Remote Observatory (AST/RO) telescope, the Steward Observatory Kitt Peak 12m telescope (KP12m) and the 10m Heinrich Hertz Telescope (HHT) between April 2001, and April 2003. We have combined 100 square arcminute maps of 0.04-0.06 pc spatial resolution and sub-km/s velocity resolution in various transitions and isotopes of CO and atomic carbon, with higher resolution mapping of the central region around IRS7 in CO(3-2) and the high density molecular tracers HCO$^+$(4-3) and H$^{13}$CO$^+$(4-3). We have also obtained a fully sampled 10\arcmin$\times$10\arcmin, 870~\micron~ continuum map of the region at 23\arcsec ~ resolution. These multi-frequency, multi-transition data give a more complete picture of the active, young and heavily embedded star formation in the R~CrA region. A summary of the observations is given in Table 1. Figure 1 shows spectra in all transitions at the position of IRS7. The spectra were generated by interpolating all the data cubes to the same number of spatial pixels (25$\times$25) and convolving them to match the 95\arcsec ~ beamsize of AST/RO at 460 GHz. 

\subsection{CO(4-3) and [CI] $^3$P$_1$-$^3$P$_0$ Mapping}

CO(4-3) and [CI] $^3$P$_1$-$^3$P$_0$ observations were performed at the 1.8m Antarctic Submillimeter Telescope and Remote Observatory (AST/RO) between November, 2001 and January, 2002 (Stark et al. 2001). For the remainder of this paper we will omit the notation for the lower level of the atomic fine structure transition for brevity. Maps were 10\arcmin$\times$10\arcmin\ (0.4pc$\times$0.4pc) in size, fully sampled and centered on IRS7. Data were taken with the facility 490/810 GHz receiver, WANDA, in the position switched mode. Normally, CO(7-6) and [CI] $^3$P$_2$ are simultaneously observed along with CO(4-3) or [CI] $^3$P$_1$, but these data were taken in the Austral summer. While weather conditions were adequate for 460 and 490 GHz observations, the atmospheric opacity was too high to effectively observe in the 850 GHz (350 \micron) atmospheric window. During observations, system temperatures ranged between 1500K and 9000K with typical values of $\sim$4000K. Facility AOS spectrometers with 1 GHz bandwidth and 1 MHz (0.6 km/s) resolution were used. Calibration was done using two actively cooled cold loads at 15K and 60K. Efficiency was measured to be $\eta_{tel} \sim 0.81$, with a beamsize of 95\arcsec\ (0.06pc) (Stark et al. 2003). Data were exported with the facility COMB software tool to individual FITS spectra. These were then compiled into a single CLASS format data file. The data were then fit with linear baselines. Processed data cubes were exported in FITS format for further processing with IDL. Typical RMS noise for the reduced data cubes is 0.19K and 0.20K for CO(4-3) and [CI] respectively. Note that isotopic observations of CO were not made. $^{13}$CO(4-3), C$^{18}$O(4-3) and C$^{17}$O(4-3) all occur at frequencies that are unobservable from the ground, as they are all obscured by pressure broadened atmospheric water lines. 

\subsection{HCO$^+$(4-3) and CO(3-2) Mapping}

In April, 2001 we made fully sampled 120\arcsec$\times$80\arcsec\ (0.07pc$\times$0.05pc) maps centered on the embedded infrared source IRS7 in $^{12}$CO(3-2), H$^{12}$CO$^+$(4-3) and H$^{13}$CO$^+$(4-3), at the HHT. Data were taken in the On-The-Fly mode with the facility dual polarization 345 GHz receiver built by the Max-Plank-Institut fur Radioastronomie in Bonn, Germany. Typical system temperatures at 15 degrees elevation were $T_{sys} \sim 2500K$. The beamsize was 23\arcsec\ (0.01pc), and the telescope efficiency was $\eta_{tel} \sim 0.7$ (for extended sources). RMS noise for the completed maps was 0.18K, 0.14K and 0.14K respectively. Backends were 62.5 kHz, 250 kHz and 1 MHz resolution filterbank spectrometers. The 250 kHz data was used for this work (0.2 km/s). Calibration was done with an ambient temperature and liquid nitrogen load, using the Hot-Sky-Cold method. Baselines were subtracted using a linear fit, and data was gridded and convolved to a 23\arcsec\ beam using the GILDAS software package. Data cubes were then exported in the FITS format for further processing using the IDL data processing package.

\subsection{$^{12}$CO(1-0), $^{13}$CO(1-0) and C$^{18}$O(1-0) Mapping}

CO(1-0) observations were performed at the Kitt Peak 12m telescope in April, 2003. The fully sampled maps are 10\arcmin$\times$10\arcmin\ (0.4pc$\times$0.4pc), centered on IRS7. Data were taken with the facility 3mm dual polarization, single-sideband receiver in On-The-Fly mode. The Millimeter-wave AutoCorrelator System (MACS) hybrid digital autocorrelation spectrometer was used as the backend, with 600 MHz bandwidth and 300 kHz (0.8 km/s) resolution. System temperatures varied between 250K and 1000K. Several OTF maps per transition were co-added to produce the final data cube. Raw data had linear baseline fits subtracted, and were gridded and convolved to a 60\arcsec\ (0.04pc) beamsize using AIPS software with the standard OTF reduction tools included with the latest AIPS release. Baseline subtracted and gridded data cubes were then exported via the FITS format for further processing with IDL. Typical RMS noise levels are 0.08K, 0.01K and 0.03K for $^{12}$CO, $^{13}$CO and C$^{18}$O respectively.

\subsection{870~\micron~ Continuum Mapping}

870~\micron~ continuum observations were made at the 10m HHT in March, 2003. Maps were 10\arcmin$\times$10\arcmin\ (0.4pc$\times$0.4pc) centered on IRS7, and fully sampled. Mapping was done with the facility 19-channel, $^3$He bolometer array used in the On-The-Fly mode. Beamsize was 23\arcsec\ (0.01pc), with a typical efficiency of $\eta_{tel} \sim 0.70$. Data consists of a single OTF map made in less than one hour. Weather was excellent, with the zenith atmospheric opacity measured at 225GHz of less than 0.05. Opacity at the observation frequency and azimuth was determined via skydips. Fits to this data show opacity at 343 GHz to be 0.260. Absolute calibration was done using Mars, assuming a flux at 343 GHz of 741.8 Jy, determined with the GILDAS ASTRO task. Reductions were performed using the GILDAS NIC package. Scans were despiked, gridded and restored using standard NIC tools. Reduced and calibrated maps were exported via the FITS format for further processing with IDL. RMS noise in the map is 0.3 Jy/beam, assuming the worst case noise at the edge of the map. 

\section{Analysis}

 \subsection{Morphology}

Our survey concentrates on a 10\arcmin$\times$10\arcmin ~ field centered around IRS7. This encompasses the peak of the sub-mm continuum emission near IRS7, and is a local maximum for the molecular gas emission. Since optical depth effects severely effect the morphology of the $^{12}$CO measurements, we use our 870~\micron~ continuum map, along with the optically thin tracers of $^{13}$CO(1-0) and C$^{18}$O(1-0) to trace the dust and gas mass in the region. Both the gas and dust show the same qualitative morphology, peaking around IRS7, with extended emission towards the northwest. The continuum emission is very strongly peaked around IRS7, with extended emission at flux levels $\sim$10 times lower than the peak. Emission in optically thin CO isotopes peaks at the same location, but is much more extended, with a second peak $\sim$200\arcsec ~ to the northwest. Emission above the 3$\sigma$ detection limit extends throughout most of the mapped region, but is generally oriented from the SE to NW. Features in the CO emission do not tend to correlate with the positions of the brightest infrared point sources in the region. A map of the central 120\arcsec$\times$80\arcsec\ in the high density tracer HCO$^+$(4-3) at 23\arcsec ~ spatial resolution reveals the ``molecular disk'' detected by \cite{and97-1}. The high density gas traces a structure elongated in Dec relative to RA with an aspect ratio of about 3:1. The continuum data, with the same spatial resolution, also clearly shows this elongated structure in the central region of the map. A hint of the structure is also seen in the CO maps, but the larger 60\arcsec ~ beamsize of these data mask the disk. Maps of these data are presented in Figure 2. 

\subsection{Far Infrared Spectral Energy Distribution}

We have combined our measured peak flux for the 870~\micron~ emission with 1.3mm data from \cite{hen94} and IRAS 100 \micron, 60 \micron, 25 \micron ~ and 12 \micron ~ fluxes from the IRAS point source catalog to produce a FIR Spectral Energy Distribution (SED) of the R~CrA core. The 1.3 mm and 870~\micron~ measurements were done via OTF mapping with virtually identical beam sizes. The IRAS beam is much larger ($\sim$100\arcsec), but the mm and sub-mm emission is strongly peaked at the central position. This large beam could cause significant contamination of these frequency points. We consider only the 100~\micron~ and 60~\micron~ fluxes when fitting a single diluted blackbody to the SED, since the two shorter wavelength bands are likely to contain contributions from hot dust. The  SED shape bears this out. Since we are only concerned with the cold dust associated with the mm and sub-mm line emission, we do not attempt to fit a hotter second blackbody component. We fit the FIR spectral energy distribution with a diluted blackbody SED, varying the parameters of source size, temperature and spectral index \citep{wal90-1}. Figure 3 shows the SED with our fit to the four lowest frequency points. We determine the fit parameters to be $\Omega_s=3.05 \times 10^{-9} Sr$, $T=36.4 K$ and $\beta=1.66$. Formal errors in the fit are insignificant compared to the uncertainties involved in the simple model for the SED and the later estimates of dust and gas masses. The temperature and spectral index are comparable to other dust cores \citep{wal90-1}, but the beam filling factor is rather large at 0.31. This is explained by the close proximity of R~CrA, only 130 pc. We then measure the FIR luminosity by integrating under the fit, and determine the luminosity to be 21.1 L$_\odot$, typical for a low mass, young protostar. We can also use these data to estimate the dust and gas mass in the core, if the dust is optically thin. The dust opacity at 870~\micron~ is 0.016, allowing us to follow Hildebrand to estimate the dust mass \citep{hil83}:

\begin{equation}
M_d=\frac{F_\nu D^2}{Q_\nu B_\nu (T)} \frac{4 a \rho}{3}
\end{equation}

\noindent
where

\begin{equation}
Q_\nu=7.5 \times 10^{-4} \left (\frac{125~\mu m}{\lambda} \right )^\beta
\end{equation}

\noindent
and {\it a} and $\rho$ are the grain radius and density. We assume 0.1 \micron ~ for {\it a} and $3~g~cm^{-2}$ for the grain density. Assuming a gas to dust ratio {\it f} of 100, we can estimate the molecular hydrogen mass and column density. We calculate a $H_2$ mass of 0.6 M$_\sun$, and a $H_2$ column density of $7.3 \times 10^{23}~cm^{-2}$ for the central 23\arcsec ~ region centered on the peak of the FIR emission. Using the dust opacities of Draine and Lee the results are lower by about 30\% \citep{dra84}. These estimates are accurate only to within a factor of several, due to the uncertainties introduced by the large IRAS beam, but are consistent with the parameters determined through other methods.

\subsection{Core Mass and Kinematics via HCO$^+$(4-3)}

Using the higher spatial resolution (23\arcsec) 120\arcsec $\times$ 80\arcsec ~ H$^{12}$CO$^+$(4-3) and H$^{13}$CO$^+$(4-3) maps of the IRS7 region, we have determined the central mass of the FIR emission peak through LTE estimates of the gas column density and through the kinematic signature of the rotation of the molecular disk. Anderson  determined the central mass using HCO$^+$(4-3) observations using the kinematic method. This gives four independent measures of the central mass of the dust core, giving us some confidence in determining the properties of the central source \citep{and97-2}. 

With both H$^{12}$CO$^+$ and H$^{13}$CO$^+$ measurements at the central position, we can estimate the column density of HCO$^+$, and then estimate a column density of $H_2$ assuming an abundance for HCO$^+$. For simplicity, we will assume the gas is in local thermodynamic equilibrium. The line shape of H$^{12}$CO$^+$ clearly shows the effects of self-absorption, so the LTE formalism will overestimate optical depth and therefore overestimate the column density and mass. Effects of self absorption on the optically thick line's integrated intensity is unlikely to be more than a factor of two, overestimating the mass by a factor of a few. More complicated microtubulent methods can be used to take the effects of self absorption into account, but many more assumptions are necessary.  We calculate the optical depth from the ratio of the optically thick to optically thin species:

\begin{equation}
\frac{I_{thick}}{I_{thin}}=\frac{1-e^{-r_a \tau}}{1-e^{-\tau}}
\end{equation}

\noindent
where $I_{thick}$ and $I_{thin}$ are the integrated intensities of the optically thick and thin species, respectively, $r_a$ is the abundance ratio of the optically thick to thin species, and $\tau$ is the optical depth of the optically thin species. For our analysis, we assume a H$^{12}$CO$^+$ to H$^{13}$CO$^+$ abundance ratio of 60 and that the gas excitation temperature is given by the peak H$^{12}$CO$^+$ line temperature divided by the telescope efficiency ($\sim$0.7). This most likely underestimates the excitation temperature, which partially counters the effect of the overestimate in optical depth. From the optical depth and excitation temperature, we can then calculate the column density of the optically thin species $N_{\nu,thin}$:

\begin{equation}
N_{\nu,thin}=T_{ex} \tau \Delta \nu \frac{6 k}{8 \pi^3 \nu \mu^2} \left (\frac{2 l +1}{2 u-1} \right ) \left ( \frac{ e^{~\frac{l}{2} \frac{h \nu}{k T_{ex}}}}{ 1 - e^{ -\frac{h \nu}{k T_{ex}} }} \right ) 
\end{equation}

\noindent
where {\it l} is the lower rotational level, {\it u} is the upper rotational level, $\mu$ is the dipole moment of the molecule and $\Delta \nu$ is the frequency interval over which the integrated intensity is measured. At the peak of the map, we calculate an excitation temperature of 15K, an optical depth for H$^{13}$CO$^+$ of 0.04, and a column density of $H_2$ of $2.2 \times 10^{23}$cm$^{-2}$. This translates to a gas mass in the central beam of 0.5 M$_\sun$, for an abundance ratio of H$^{12}$CO$^+$ to $H_2$ of $10^{-9}$ \citep{hog03}.

We can also estimate core mass through gas kinematics from velocity centroid maps. We calculate the velocity centroid of each map position following \cite{ade82}:

\begin{equation}
v_c=\frac{\sum_i (V_{Dopp})_i (T_A)_i}{\sum_i (T_A)_i}
\end{equation}

\noindent
where $V_{Dopp}$ is the velocity offset from line center and $T_A$ is the line brightness at the offset velocity, indexed by velocity channel {\it i}. The result is a map of line centroid velocity, where the line centroid is the velocity at which the integrated intensity redward and blueward of the centroid velocity is the same. This gives a measure of the kinematics of the object averaged over the line. We compute the centroid velocity over the range -3.0 to +2.5 km/s, which encompasses the entire line core at the central position, and averages over self-absorption effects. The resulting map is shown in Figure 4. We take a slice at a position angle of 120 degrees, orthogonal to the velocity gradient. The map has been interpolated by a factor of 3 in both dimensions. A slice centered on IRS7 was averaged over three columns to produce the rotation curve. We then assume circular motion around a point mass to determine an upper limit for the central mass:

\begin{equation}
M_c=\frac{v^2 r}{G}
\end{equation}

\noindent
where we define {\it v} as the half-amplitude of the rotation curve, and {\it r} as half the distance between peaks in the curve. We determine a rotational velocity of 0.47 $\pm$ 0.04 km/s, a radius of 0.015 $\pm$ 0.001 pc and a derived mass of 0.75 $\pm$ 0.15 M$_\sun$, assuming a 3\arcsec ~ pointing error, and errors in the velocity centroid according to \cite{nar00}. This is consistent with the result of Anderson et al. of 0.8 M$_\sun$.

\subsection{Outflow}

In the convolved spectra of Figure 1, it is clear that the optically thick species show extended wing emission at high velocities. We made outflow maps by calculating the integrated intensity in the line wings for $^{12}$CO(1-0), $^{13}$CO(1-0), $^{12}$CO(4-3) and [CI] $^3$P$_1$. We use the convolved spectra to determine the velocity range of the wing material. Figure 5 shows the resulting outflow maps, and the location of known infrared sources in the area. Optical depth effects alter the morphology of the $^{12}$CO maps as compared to the $^{13}$CO map. In [CI], only the red outflow lobe is clearly visible.

Over the velocity range -4 to 3 km/s and 8 to 15 km/s, we can use both the $^{12}$CO(1-0) and $^{13}$CO(1-0) outflow data to compute the column density of CO. The peak antenna temperature for CO(1-0), CO(3-2) and CO(4-3) are all similar (see Figure 1), suggesting the gas is thermalized. We assume the peak line temperature, corrected for efficiency, is the excitation temperature in the outflow. In addition, the abundance of CO/$H_2$, while not exactly known, is expected to vary less across the galaxy than the HCO$^+$/$H_2$ abundance ratio \citep{hog03}. As is clear from the $^{12}$CO(1-0) spectrum at the center position, the optical depth in the blue wing is much higher than the red, due to either foreground obscuration or self-absorption. We find blue wing and red wing masses of 0.46 M$_\sun$ and 0.38 M$_\sun$, respectively, assuming an abundance ratio of $^{12}$CO to $^{13}$CO of 60 and a $^{13}$CO/$H_2$ abundance ratio of $2.5 \times 10^{-6}$. The characteristic age of the outflow is found by dividing the extent of the outflow by the maximum observed velocity from line center, $v_c$. We take 300\arcsec ~ as the outflow extent, and 17 km/s as the maximum velocity, which gives a characteristic age of $10^4$ years. Using these values, we can calculate the characteristic values for the momentum, $Mv_c$, the kinetic energy, $\frac{1}{2}Mv_c^2$, the outflow mechanical luminosity, $KE/Age$, and the mass outflow rate, $M/Age$. Outflow mass and energetics are summarized in Table 2. 

All the isotopic forms of CO(4-3) are not observable from the ground because of atmospheric absorption due to water and molecular oxygen. For this transition, outflow mass and energetics are determined by assuming the line wing emission is optically thin. This assumption provides a lower limit to the outflow mass. We again assume the excitation temperature is the peak CO line temperature divided by the telescope efficiency, 0.81 for AST/RO at 460 GHz \citep{sta01}. We follow Snell to determine the column density of $^{12}$CO assuming the emission is optically thin \citep{sne84}:

\begin{equation}
N_{CO}=\frac{4.2 \times 10^{13}~T_{ex} \int T_R (CO) dV}{exp(-\frac{h \nu}{E})}~cm^{-2}
\end{equation}

\noindent
The derived outflow masses (0.40M$_\odot$ \& 0.32M$_\odot$) are similar to those determined from CO(1-0). This is not surprising if the gas in the region is relatively warm. We can also apply the same approach to the $^{12}$CO(1-0) data to see the effects of taking optical depth into account. Masses are calculated to be 0.21 M$_\sun$ and 0.06 M$_\sun$ for the red and blue wings respectively. The value for the red wing is within a factor of two of the value obtained by the LTE analysis, but the blue wing mass is several times lower. This is due to the higher optical depth in  the blue wing of the $^{12}$CO(1-0) lines. 

\section{Discussion}

\subsection{A Possible Class 0 Source Driving the FIR Emission}

In most previous research, it is assumed that the molecular line and FIR emission in the region is directly associated with R~CrA. However, work by Harju and Anderson suggested the embedded Class I source IRS7 as the most likely driving source for this emission (\cite{har93}, \cite{and97-1}, \cite{and97-2}). The possibility exists that the peak of the mm-wave and sub-mm wave emission, and the driving source for the molecular outflow(s) in the region might not be IRS7, but a more deeply embedded source not detected in infrared surveys to date. The results presented here suggest that the source of the molecular line and cold dust emission may be a deeply embedded Class 0 source within $\sim$10\arcsec\ of IRS7, with a mass of about 0.5 M$_\sun$. 

Recent work by \cite{chi03} concludes that IRS7 is not the driving source behind the mm-wave emission in the region. Their 1.3mm continuum map of the region identifies 25 dust emission peaks. The peak near IRS7, called MMS13, is conincident with neither the thermal IR source associated with IRS7 from \cite{wil97} nor the VLA sources from \cite{bro87}, but is located $\sim$15\arcsec\ south.  They conclude that MMS13 is most likely a deeply embedded Class 0 source, although their continuum flux suggests a source mass of 5 M$_\odot$ for 20K dust. Our data also seem to suggest the driving source is south of IRS7, but our spatial resolution is not adequate to make a firm determination. Recently, 50 GHz continuum interferometric imaging of the region was performed by \cite{cho04}. They find an elongated structure with a position angle of $\sim$120 degrees located $\sim$5\arcsec\ to the north of IRS7. The elongated structure has a SED consistent with dust emission, while the emission peak at the position of IRS7 has an SED suggesting free-free or non-thermal emission. Millimeter wave interferometric observations toward IRS 7 in one or more molecular lines (e.g. CO, HCO$^+$ and CS) are needed to identify the location of the driving source. 

Other evidence exists that suggests the source of the emission is a Class 0 object. The mass of the molecular core can be estimated three ways: 1) a kinematic measure (Figure 4) of the total mass of the protostar and surrounding gas. 2) the FIR continuum flux and 3) the column density of HCO$^+$. All three of these measurements have considerable uncertainty, but taken together point towards a deeply embedded source of about 0.5 M$_\sun$ with a gaseous envelope of similar mass. 

The kinematic measurement from the centroid velocity of HCO$^+$(4-3) shows a velocity gradient strongly suggestive of rotation. If this velocity gradient is due to rotation, then a lower limit to the enclosed mass can be derived. In the 0.030 pc diameter volume about the center of motion, we derive a gravitational mass of 0.75 M$_\sun$. This assumes all the mass is concentrated at the center. 

We also estimated the gas mass in the central 22\arcsec\ (about half the distance between velocity peaks in the velocity gradient) via continuum observations. Assuming the dust emission arises from 0.1 \micron\ radius silicate dust grains with a dust emissivity $\propto \lambda^{-2}$ and the dust to gas ratio is $\sim$100, we compute a total gas mass for the core of 0.6 M$_\odot$. This determination has considerable uncertainy due to the use of IRAS data in the determination of the SED, but is consistent with the masses derived using other methods.

The measurement of the $H_2$ mass via HCO$^+$ also gives a similar answer. We used H$^{12}$CO$^+$ and H$^{13}$CO$^+$ measurements with a 22\arcsec ~ beamsize to estimate the column density of $H_2$. Both the varying abundance of HCO$^+$ to $H_2$ and the effects of self absorption are uncertain. The presence of self-absorption will lead to overestimates of the HCO$^+$ column density. Assuming a HCO$^+$/H$_2$ abundance ratio of 10$^{-9}$ and neglecting the effects of self absorption, we estimate a gas mass of 0.5 M$_\sun$. While uncertain, both these measures point to a total envelope mass on the order of the mass of the central source. In addition, the somewhat small FIR luminosity of 21 L$_\odot$ is similar to low mass protostars of about 0.5 M$_\sun$ \citep{wal90-2}. We suggest that the driving source for the outflow is a previously unidentified Class 0 protostar located $\sim$ 10\arcsec\ from IRS~7 at the peak of the mm-wave continuum emission, referred to by \cite{chi03} as MMS13.

\subsection{Infall}

Spectral lines of Class 0 protostellar objects sometimes show the classic spectral signature of infall \citep{wal86}. As seen in Figure 6, HCO$^+$(4-3) spectra near the center position of the map show an enhanced blue peak, characteristic of this infall signature. In addition, the centroid velocity map in Figure 4 shows the ``Blue Bulge'' signature of infall motion \citep{wal94}. North and south of the central position, rotation begins to dominate the kinematics. At about 30\arcsec ~ east and west of the source the outflow dominates the appearance of the line profiles in the HCO$^+$ map. In CO maps, the outflow completely masks all signs of rotation and infall. Only in dense gas tracers (e.g. HCO$^+$) do all the intertwined motions reveal themselves near the source. Convolving spectra to larger beamsizes, we continue to see a line shape suggestive of infall. In Figure 7, the HCO$^+$ spectrum has been convolved in half beam increments from 22\arcsec ~ to 77\arcsec, which is essentially the entire HCO$^+$(4-3) map. The infall lineshape is still clearly visible in all the spectra, suggesting that infall is occurring over the entire central region around the protostellar source. This points toward an infall region of at least 6000 AU in radius. Using our measurements of FIR luminosity and temperature, we can estimate the mass infall rate for this object. If all the FIR luminosity is generated through accretion, then

\begin{equation}
L_{FIR}=\frac{G M \dot{M}}{r}
\end{equation}

\noindent
where {\it r} is the radius of the protostar and {\it M} is the mass of the protostar. We assume $2 \times 10^{11}$ cm for the radius, about 3 R$_\odot$ \citep{sta80}, and 0.5 M$_\sun$ for the mass of the central object, given our mass estimates of the source+envelope and envelope masses. This leads to a mass accretion rate of $4 \times 10^{-6}$ M$_\sun$/yr, uncertain to within a factor of a few. A mass infall rate can also be computer for the case of self-similar collapse.  Assuming a soundspeed due to only thermal pressure, and ignoring magnetic and turbulent support, the mass accretion rate is given by \citep{shu77}:

\begin{equation}
\dot{M}=\frac{0.975~a^3}{G}
\end{equation}

\noindent
where,

\begin{equation}
a=\sqrt{\frac{k T}{m_{H_2}}}
\end{equation}

\noindent
Substituting the gas temperature derived from the observed spectral energy distribution ($\sim$36K), we find a mass accretion rate of $1 \times 10^{-5}$ M$_\sun$/yr. Both these estimates are consistent with a low mass, Class 0 source.

\subsection{Atomic Carbon Distribution}

In the canonical picture of carbon in a molecular cloud, an "onion-skin" model is used where the carbon is ionized in the outer layer of the cloud as [CII]. Further in the cloud where the carbon is more shielded, it is in atomic form as [CI]. Deeper in the cloud, the carbon forms CO. In some cases, where the cloud material is very clumpy, this same picture holds, but now each clump acts as a  small cloud, causing [CII] and [CI] to appear well mixed throughout the cloud \citep{stu88}. We see evidence for both these effects in our carbon observations. To reveal enhancement of [CI] relative to C$^{18}$O, we calculated the column density of both species, assuming they are both optically thin. We estimated the column density of the atomic carbon following \cite{wal93-2}:

\begin{equation}
N_C=\frac{N_1}{3} \left ( e^{\frac{23.6}{T_{ex}}}+3+5 e^{\frac{-38.8}{T_{ex}}} \right )
\end{equation}

\noindent
where

\begin{equation}
N_1=5.94 \times 10^{15} \int T_{mb}~dV~cm^{-2}
\end{equation}

\noindent
We then plotted the ratio of [CI]/C$^{18}$O column density normalized to the average over the map to look for enhancement/depletion of atomic carbon relative to CO, considering only locations were both the [CI] and CO integrated intensities were larger than the 2$\sigma$ noise level. The results are shown in Figure 8. As expected from the canonical picture, [CI] is enhanced at the edges of the cloud, and depleted towards the center. We have plotted the contours over an optical image of the region from the Palomar Sky Survey. The regions of low relative [CI] abundance correlate with the regions of high optical extinction, and the region with strong FIR continuum emission.

In the outflow map of Figure 5, we see that carbon is observed throughout the red wing of the outflow. Only a small amount of carbon is visible in the blue wing. While the [CI] is not as ubiquitous as $^{13}$CO in the outflow, it is still visible, and does not appear to be in a shell configuration. This leads us to believe that while atomic carbon is enhanced at the edge of the cloud, some atomic carbon is still mixed throughout the cloud, pointing to clumpy structure even in the outflow.  

\subsection{System Configuration}

The kinematic signs of rotation, infall and outflow, combined with the distribution of CO, HCO$^+$ and submillimeter continuum emission all combine to produce a picture of star formation in the R~CrA molecular core. Due to the proximity of R~CrA, the kinematic signatures of infall, rotation and outflow can be tentatively identified and disentangled. A proposed configuration for the MMS13 core is shown in Figure 9, overlaid on the centroid velocity plot of HCO$^+$(4-3). We believe a highly embedded Class 0 protostellar source is the driver for the FIR emission in the area. Kinematic and morphological evidence point to a location within $\sim$10\arcsec\ of the Class I source IRS7. This source, located at the peak of the mm-wave continuum emission, was tenatively identified by \cite{chi03} as MMS13. This source is surrounded by a rotating molecular disk, first observed by Anderson et al. (\cite{and97-1}, \cite{and97-2}). Our HCO$^+$(4-3) measurements point to a similar conclusion, with an enclosed mass of 0.75 M$_\sun$ $\pm$ 0.15 M$_\sun$. Mass estimates based on derived gas column density, 870 \micron\ continuum measurements and dynamical arguments suggest a central protostar of $\sim$0.5 M$_\sun$, with an envelope of similar mass. We postulate the disk major axis is at a position angle of about 60 degrees. This position angle gives the cleanest rotational signature in the velocity gradient, and is consistent with the emission and the orientation of the molecular outflow. This molecular disk has a major axis of $\sim$2500 AU, and is unresolved in the orthogonal direction in all our observations. In the central 20\arcsec-30\arcsec ~ of the HCO$^+$(4-3) map, the line profiles show a strong infall line shape. This shape persists when the spectra are convolved to beamsizes as large as the entire HCO$^+$ map. In addition, the central region of the centroid velocity plot shows the "Blue Bulge" signature of infall motion. This infall region is at least 6000 AU in radius, surrounding the molecular disk. At distances larger that 6000 AU, the kinematics of the molecular outflow begin to dominate the line profiles. At lower spatial resolution, and with more sensitivity to the lower density gas, CO maps are virtually all dominated by the outflow and the ambient cloud material.

\section{Summary}

We have observed the molecular core near R~CrA in $^{12}$CO(1-0), $^{13}$CO(1-0), C$^{18}$O(1-0), $^{12}$CO(3-2), $^{12}$CO(4-3), [CI] $^3$P$_1$-$^3$P$_0$, H$^{12}$CO$^+$(4-3), H$^{13}$CO$^+$(4-3) and 870~\micron~ continuum emission. These data suggest that the source for the FIR line and continuum emission, and the driving source for the molecular outflow is a yet unidentified Class 0 source deeply embedded in the cloud. Using HCO$^+$ and continuum observations, we derive a core gas mass of about 0.5 M$_\sun$. From a centroid velocity analysis of the HCO$^+$ emission, we estimate the mass enclosed inside the rotating molecular disk to be $\sim$0.75 M$_\sun$. Both these measurements suggest the presence of a low mass (0.5 M$_\sun$) protostar surrounded by an envelope of similar mass. 

Lineshapes show evidence for infall, rotation and outflow motions in the central 60\arcsec $\times$ 60\arcsec ~ of the map. The outflow dominates the line formation outside this region. The outflow mass is $\geq~$0.84 M$_\sun$, similar to the mass of the protostar we believe is driving the outflow. The energetics are typical for a molecular outflow around a young, low mass protostar, with a mechanical luminosity much less than the FIR luminosity.

Our result supports the "standard" model for low mass star formation. Follow-up with mm-wave interferometers in both continuum and high density molecular tracers could reveal the Class 0 source, if present, and allow detailed study of the infall and rotational motions of the molecular disk.  

\acknowledgments
We thank the staff of the Arizona Radio Observatory, particularly Harold Butner for collecting the 870\micron\ continuum data during priority observing mode. We also thank the staff of the Antarctic Submillimeter Telescope and Remote Observatory for their support of the AST/RO telescope. Michael Meyer, John Bieging, and Erick Young also participated in valuable discussions. This research was supported in part by the National Science Foundation under a cooperative agreement with the Center for Astrophysical Research in Antarctica (CARA), grant number NSF OPP 89-20223. Author C. Groppi thanks the National Science Foundation (grant number 0138318) and the National Aeronautics and Space Administration (grant number S01-GSRP-023) for their support. The National Radio Astronomy Observatory is a facility of the National Science Foundation operated under cooperative agreement by Associated Universities, Inc.

\clearpage

\figcaption[fig1.eps]{Spectra centered on IRS7 from all our data sources. Spectra were generated by re-gridding all the maps to the same number of pixels, then by convolving the maps to match the largest beamsize, 95\arcsec. Optically thick CO transitions are dominated by outflow kinematics, and are subject to significant self-absorption. \label{fig1}}

\figcaption[fig2.eps]{10\arcmin $\times$ 10\arcmin ~ maps in 870~\micron~ continuum emission, $^{13}$CO integrated intensity, C$^{18}$O integrated intensity, and a 80\arcsec $\times$ 120\arcsec ~ map in HCO$^+$(4-3) integrated intensity. Minimum contours are at the 2$\sigma$ level, with 2$\sigma$ contour spacing. Beamsizes are specified in each map. The continuum emission from cold dust is well traced by the optically thin CO emission. The HCO$^+$ inset shows the ''molecular disk'' structure reported by \cite{and97-2}, elongated along the N-S axis. \label{fig2}}

\figcaption[fig3.eps]{The FIR SED of the R~CrA region centered on IRS7. The 870~\micron~ point is from this work. The 1300~\micron~ point is from \cite{hen94}, data taken with the 19 channel facility bolometer array at SEST. The FIR data are IRAS measurements of R~CrA. Since the IRAS beam is $\sim$100\arcsec, some contamination from the surrounding material off the dust emission peaks is present. We fit only the 100~\micron~ and 60~\micron~ points to exclude contributions from hot dust. Error bars are $\pm$ 1$\sigma$. \label{fig3}}

\figcaption[fig4.eps]{HCO$^+$(4-3) centroid velocity plot. Contours are 0.25 km/s, with dotted contours denoting negative velocity. Below is a slice denoted by the box showing the signature of rotation. The centroid velocity plot also shows evidence of the ''Blue Bulge,'' a sign of infall motion. \label{fig4}}

\clearpage

\figcaption[fig5.eps]{Outflow maps of R~CrA. Panels are $^{12}$CO(1-0), $^{12}$CO(4-3), $^{13}$CO(1-0) and [CI] $^3$P$_1$. Minimum contours are 2$\sigma$, with 2$\sigma$ contour spacing. The $^{12}$CO(1-0) and $^{12}$CO(4-3) blue wing velocity interval is -10 to 3 km/s, while the red interval is 8 to 22 km/s. The $^{13}$CO(1-0) and [CI] blue wing interval is -4 to 3 km/s, while the red wing interval is 8 to 15 km/s. Dashed contours denote negative velocities. \label{fig5}}

\figcaption[fig6.eps]{H$^{12}$CO$^+$(4-3) spectral line mosaic of a 60''$\times$60'' region centered on IRS~7. The y-axis of each spectrum is $T_A^*$(K) and the x-axis is velocity (km/s). The box in figure 2 marks the location of the mosaic. \label{fig6}}

\figcaption[fig7.eps]{HCO$^+$(4-3) spectra, centered on IRS7. Spectra were produced from an OTF map, convolved in half beam increments from 22\arcsec to 77\arcsec. \label{fig7}}

\figcaption[fig8.eps]{The normalized [CI] $^3$P$_1$ to C$^{18}$O(1-0) column density ratio. Contours are overlaid on a Palomar Sky Survey image of the region. In areas of high visual extinction, the abundance of [CI] drops relative to CO as expected, with and increased [CI]/CO ratio at the edges of the regions of high visual extinction. \label{fig8}}

\figcaption[fig9.eps]{A schematic of our proposed system configuration. The contours show the combined effects of rotation, infall and outflow. The morphology suggests the existence of a Class 0 source within $\sim$10\arcsec\ of IRS7, surrounded by a $\sim$2500 AU radius molecular disk at a position angle of 60 degrees. The envelope surrounding this disk exhibits signs of infall. Toward the east and west edges of the map, outflow kinematics begin to dominate. The diagram is overlaid on the HCO$^+$(4-3) centroid velocity plot. \label{fig9}}

\clearpage

\begin{deluxetable}{cccccc} 
\tabletypesize{\scriptsize}
\tablecaption{R CrA Outflow Energetics \label{tab2} }
\tablewidth{0pt}
\tablecolumns{6}
\tablehead{ 
\colhead{Line Wing} & 
\colhead{Mass} & 
\colhead{Momentum} & 
\colhead{Kinetic Energy} & 
\colhead{Mechanical Luminosity} & 
\colhead{Mass Outflow Rate} \cr
\colhead{\& Method} &
\colhead{M$_\odot$} &
\colhead{$g~cm~s^{-1}$} &
\colhead{$g~cm^2s^{-2}$} &
\colhead{L$_\odot$} &
\colhead{M$_\odot~yr^{-1}$} 
}
\startdata
CO(1-0) LTE red & 0.38 & 1.3$\times 10^{39}$ & 1.1$\times 10^{45}$ & 0.44 & 3.6$\times 10^{-5}$ \\ 
CO(1-0) LTE blue & 0.46 & 1.6$\times 10^{39}$ & 1.3$\times 10^{45}$ & 0.53 & 4.3$\times 10^{-5}$ \\ 
$^{12}$CO(1-0) Optically thin red & 0.21 & 7.2$\times 10^{38}$ & 6.1$\times 10^{44}$ & 0.25 & 2.0$\times 10^{-5}$\\ 
$^{12}$CO(1-0) Optically thin blue & 0.06 & 2.2$\times 10^{38}$ & 1.9$\times 10^{44}$ & 0.07 & 6.0 $\times 10^{-6}$\\ 
$^{12}$CO(4-3) Optically thin red & 0.40 & 1.4$\times 10^{39}$ & 1.2$\times 10^{45}$ & 0.46 & 3.8$\times 10^{-5}$ \\ 
$^{12}$CO(4-3) Optically thin blue & 0.32 & 1.1$\times 10^{39}$ & 9.3$\times 10^{44}$ & 0.37 & 3.1$\times 10^{-5}$ \\ 
\enddata
\end{deluxetable}

\end{document}